\documentclass[aps,twocolumn,showpacs,superscriptaddress,amsmath,amssymb,amsfonts,floatfix,longbibliography]{revtex4-2}

\usepackage[T1]{fontenc} 
\usepackage{graphicx}
\usepackage{float}
\usepackage{dcolumn}
\usepackage{bm}
\usepackage{amssymb}
\usepackage{tabularx}
\usepackage{microtype}
\usepackage{xfrac}
\usepackage{chemformula}
\usepackage{gensymb}
\usepackage[most]{tcolorbox}
\usepackage{xcolor}
\usepackage{multirow}
\usepackage{enumitem}
\usepackage{array}
\usepackage{booktabs}
\usepackage[flushleft]{threeparttable}
\usepackage{natbib,hyperref}
\usepackage{graphicx}
\usepackage{makecell}
\usepackage{colortbl}
\usepackage{xcolor}
\usepackage{soul}

\hypersetup{
	colorlinks=true, 
	citecolor=blue,  
}

\begin{document}

\title{The origin and scarcity of breathing pyrochlore lattices in spinel oxides}

\author{Valentina Mazzotti}
    \affiliation{Stewart Blusson Quantum Matter Institute, University of British Columbia, Vancouver, BC V6T 1Z4, Canada}
    \affiliation{Department of Physics \& Astronomy, University of British Columbia, Vancouver, BC V6T 1Z1, Canada}

\author{Solveig S. Aamlid}
    \affiliation{Stewart Blusson Quantum Matter Institute, University of British Columbia, Vancouver, BC V6T 1Z4, Canada}

\author{Abraham A. Mancilla}
      \affiliation{Stewart Blusson Quantum Matter Institute, University of British Columbia, Vancouver, BC V6T 1Z4, Canada}
    \affiliation{Department of Physics \& Astronomy, University of British Columbia, Vancouver, BC V6T 1Z1, Canada}

\author{Janna Machts}
    \affiliation{TRIUMF, Vancouver, British Columbia, V6T 2A3 Canada}
    \affiliation{School of Physics and Astronomy, University of Edinburgh, Edinburgh EH9 3FD, United Kingdom}

\author{Megan Rutherford}
    \affiliation{Stewart Blusson Quantum Matter Institute, University of British Columbia, Vancouver, BC V6T 1Z4, Canada}
    \affiliation{Department of Physics \& Astronomy, University of British Columbia, Vancouver, BC V6T 1Z1, Canada}    

\author{J\"org Rottler}
    \affiliation{Stewart Blusson Quantum Matter Institute, University of British Columbia, Vancouver, BC V6T 1Z4, Canada}
    \affiliation{Department of Physics \& Astronomy, University of British Columbia, Vancouver, BC V6T 1Z1, Canada}
    
\author{Kenji M. Kojima}
    \affiliation{TRIUMF, Vancouver, British Columbia, V6T 2A3 Canada}
    \affiliation{Stewart Blusson Quantum Matter Institute, University of British Columbia, Vancouver, BC V6T 1Z4, Canada}
    
\author{Alannah M. Hallas}
\email[Email: ]{alannah.hallas@ubc.ca}
    \affiliation{Stewart Blusson Quantum Matter Institute, University of British Columbia, Vancouver, BC V6T 1Z4, Canada}
    \affiliation{Department of Physics \& Astronomy, University of British Columbia, Vancouver, BC V6T 1Z1, Canada}
    \affiliation{Canadian Institute for Advanced Research (CIFAR), Toronto, ON, M5G 1M1, Canada}

\date{\today}
\begin{abstract}
Breathing pyrochlores are a unique class of materials characterized by a three-dimensional lattice of corner-sharing tetrahedra. 
However, unlike conventional pyrochlores where all tetrahedra are identical in size, the breathing pyrochlore lattice is composed of alternating large and small tetrahedra. Experimental realizations of the breathing pyrochlore lattice are rare but they do occur in $A$-site ordered spinels, as in the prototype materials LiGaCr$_4$O$_8$ and LiInCr$_4$O$_8$. \textcolor{black}{In this work, we demonstrate that Cr cannot be straightforwardly substituted with other magnetic transition metals while retaining the breathing pyrochlore structure.} To explain this observation, we perform density functional theory (DFT) calculations to investigate the formation and stability of LiGaCr$_4$O$_8$ and LiInCr$_4$O$_8$, focusing on the energy scales associated with $A$ and \(B\)-site orderings as well as the magnetic exchange interactions of Cr ions. We identify the strong octahedral site preference of \ch{Cr^{3+}} as a key factor in protecting the structural integrity of the pyrochlore sublattice, which in turn enables the breathing distortion to proceed. We also demonstrate that the charge order between the $A$ (Li) and $A'$ sites (Ga or In) is maintained at all temperatures up to decomposition. Furthermore, while the magnetic exchange interactions constitute a relatively small energy scale and therefore do not play a fundamental role in the structural stability of LiGaCr$_4$O$_8$ and LiInCr$_4$O$_8$, magnetism may play a critical role in setting the magnitude of the tetrahedral distortion. Ultimately, we conclude that the scarcity of breathing pyrochlores is a consequence of the stringent requirement on site ordering for their formation.
\end{abstract}

\maketitle

\section*{\label{sec:Introduction}Introduction}
Complex oxides display remarkable functionalities that are intricately linked with their cation orderings. \textcolor{black}{Isomorphic complex oxides often form solid solutions where some cations share a single sublattice, depending on considerations related to charge, ionic radii, and polarizability~\cite{Goldschmidt1926,Pauling1929, Hume-Rothery1935}. In the opposing limit, complex oxides sometimes form superstructures in which the cations order on distinct sublattices, as observed in double perovskites~\cite{anderson1993b,vasala2015a2b} and rocksalt-type lithium transition metal oxides~\cite{Clement2020}}. Consequently, the tendency for oxides to tolerate disorder and their propensity to realize more complex ordered structures are key factors in optimizing their functionalities.

\begin{figure*}[htbp]
  \centering
    \includegraphics[width=\linewidth]{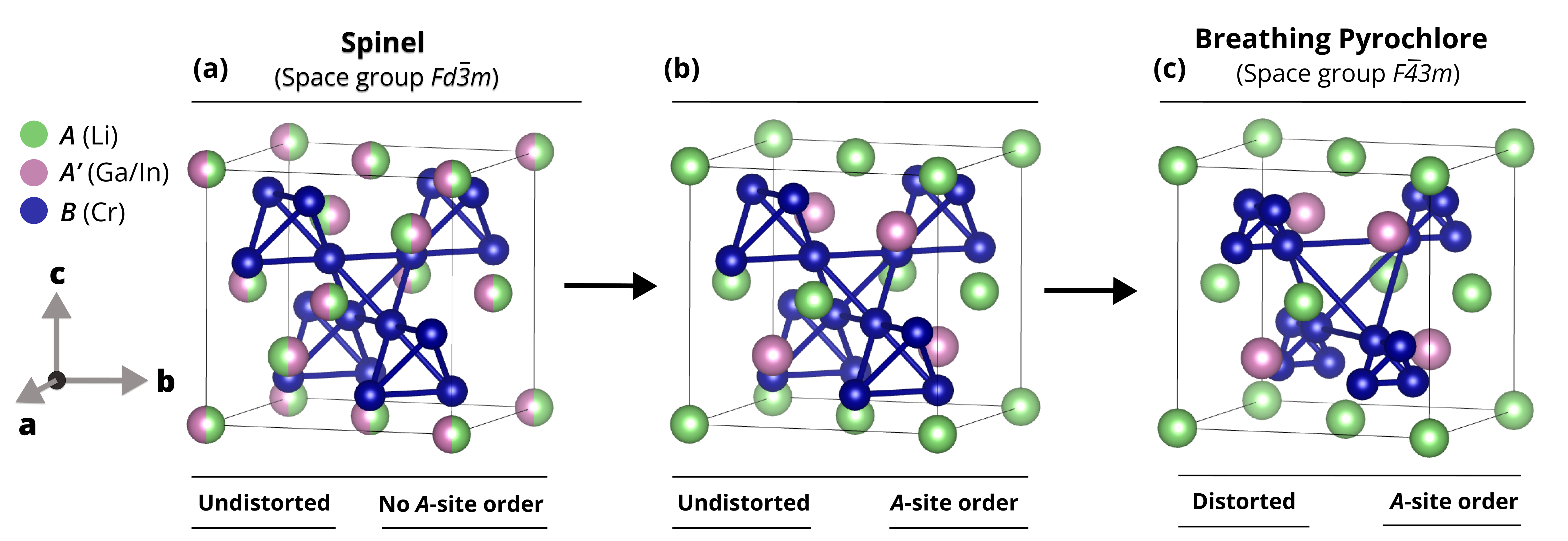}
  \caption{\textbf{The relationship between the cation disordered spinel and the distorted breathing pyrochlore lattice.} \textbf{(a)} The parent phase for the breathing pyrochlores, LiGaCr$_4$O$_8$ and LiInCr$_4$O$_8$, is a cubic spinel (space group $Fd\overline{3}m$ with chemical formula $AB_2$O$_4$ in which the cations sharing the $A$-site are randomly distributed and the $B$-site cation forms an undistorted corner-sharing tetrahedral (pyrochlore) lattice. \textbf{(b)} We show that cation ordering of the $A$-site energetically precedes the distortion of the pyrochlore sublattice, \textcolor{black}{as represented by this hypothetical intermediate structure, which can be interpreted as a site-ordered but undistorted spinel}. \textbf{(c)} The experimentally observed breathing pyrochlore structure for LiGaCr$_4$O$_8$ and LiInCr$_4$O$_8$, in the cubic $F\overline{4}3m$ space group with chemical formula $AA'B_4$O$_8$. This structure exhibits both $A$-site cation ordering and a distorted breathing pyrochlore $B$-site sublattice characterized by alternating small and large tetrahedra, which have been exaggerated here for visual clarity.}
  \label{fig:Fig1-SpinelToBreathingPyrochlore}
\end{figure*}

In this context, one particularly interesting family of materials is the spinel oxides, with the chemical formula $AB_2$O$_4$, shown in Fig~\ref{fig:Fig1-SpinelToBreathingPyrochlore}(a). The spinel structure is made up of tetrahedrally oxygen-coordinated cations at the $A$ site and octahedrally oxygen-coordinated cations at the $B$ site, which form diamond and pyrochlore lattices, respectively. Spinels are known for their chemical versatility, \textcolor{black}{their ability to form solid solutions~\cite{hill1979systematics,sickafus1999structure,Navrotsky1969}, and their}
 structural instabilities that involve charge, spin, and orbital degrees of freedom~\cite{goodenough1955theory,lee2000local,wheeler2010spin,senn2012charge}. One well understood structural phenomenon in the spinels is their tendency towards cation inversion~\cite{verwey1947physical,o1983simple}. In structures termed normal spinels, the $A$ site is exclusively occupied by a divalent cation, and the $B$ site is occupied by trivalent cations. In partially or fully inverse spinels, the $A$ site is partially or completely occupied by a trivalent cation. The degree of inversion is often dictated by crystal field effects and can, in some cases, be experimentally controlled through synthesis conditions~\cite{schmocker1976inversion,o1991temperature}. \textcolor{black}{In the fully inverted limit, or in cases where two or more distinct cations share the octahedral site, spinels will often undergo cation ordering on the $B$-sublattice}, leading to more complex structural motifs~\cite{liu2019unified,pilania2020prediction,kocevski2020high}. By comparison, cation order on the $A$ sublattice is rare, with the notable exception of the ``breathing pyrochlore’’ distortion, which is the focus of this work.

Spinels with a breathing pyrochlore distortion crystallize in the space group \( F\overline{4}3m \), which is a subgroup of the conventional spinel \( Fd\overline{3}m \) \cite{Joubert1966,talanov2020formation}. Charge ordering of two cations that share the $A$-site leads to compounds with the formula unit $AA'B_4$O$_8$, which can be understood as a doubling of the standard spinel formula unit $AB_2$O$_4$~\cite{BreathingPyrochlores3}. The monovalent $A$ and trivalent $A’$ cations occupy alternating tetrahedral sites producing a relative contraction and expansion of their local environments. This ordering of the $A$ site cation in turn disrupts the $B$ cation sublattice, which in conventional spinels forms a uniform pyrochlore lattice. The resulting breathing pyrochlore sublattice of the $B$ site cations is characterized by a corner-sharing array of alternating small and large tetrahedra \cite{Lee2016}, as can be seen in Fig.~\ref{fig:Fig1-SpinelToBreathingPyrochlore}(c). \textcolor{black}{The degree of the tetrahedral distortion can be quantified by the ratio of the bond distances, $d/d'$, where $d$ and $d'$ are the cation-cation bond distances within the small and large tetrahedra, respectively.} Alternatively, in cases where the breathing pyrochlore lattice is occupied by a magnetic cation, the \textcolor{black}{breathing} factor can be quantified by the ratio of the magnetic exchange couplings, $B_f  = J'/J$ where $J$ and $J'$ are the coupling parameters in the small and large tetrahedra, respectively.

Breathing pyrochlore lattices have attracted a lot of interest in the context of their complex magnetism and geometric frustration~\cite{Okamoto2013,tanaka2014novel,nilsen2015complex,li2016weyl,Ghosh2019,yan2020rank,he2021neutron}. However, a limiting factor has been the scarcity of materials realizations. There are only five known $A$-site ordered spinels and of these, two are nonmagnetic (LiGaRh$_4$O$_8$ and LiInRh$_4$O$_8$~\cite{Joubert1966}) and a third has a secondary magnetic element off the breathing pyrochlore sublattice complicating its description (LiFeCr$_4$O$_8$~\cite{saha2017magnetostructural}). The two remaining magnetically unambiguous examples are LiGaCr$_4$O$_8$ and LiInCr$_4$O$_8$, which both have Cr$^{3+}$ $S=\sfrac{3}{2}$ occupying the breathing pyrochlore $B$ sublattice~\cite{Joubert1966,Okamoto2013}. Beyond spinels, one other notable magnetic breathing pyrochlore is Ba$_3$Yb$_2$Zn$_5$O$_{11}$, where the extremely large breathing factor results in decoupled small tetrahedra~\cite{kimura2014experimental,rau2016anisotropic,haku2016low,dissanayake2022towards}. To explore the range of possible magnetic states, $A$-site ordered spinels therefore remain the best option. The limited number of known breathing pyrochlore spinels stands in stark contrast to undistorted spinels, which are known for their ability to accommodate a large swath of the periodic table, raising questions about the factors determining their structural stability. 

In this paper, we aim to address the scarcity of spinels with breathing pyrochlore distortions by quantifying the relevant energy scales dictating the stability of LiGaCr$_4$O$_8$ and LiInCr$_4$O$_8$ and those that preclude the chemical substitution of Cr with other magnetic transition metals. We first explore the energy scale associated with $A$-site ordering, finding that the charge order of the $A$ (Li) and $A'$ (Ga or In) cations is energetically favored well above the synthesis temperature and the resulting electrostatic effects drive the formation of the breathing pyrochlore lattice. We next examine the energetics of the $B$-site, revealing that the strong octahedral site preference of Cr$^{3+}$ is needed to preserve the site ordering of the $B$-site and is a prerequisite to obtaining a breathing pyrochlore distortion. Finally, we extract the magnetic exchange interactions in LiGaCr$_4$O$_8$ and LiInCr$_4$O$_8$ and demonstrate that the magnetic couplings play an important role in setting the magnitude of the breathing pyrochlore distortion although the energies involved are comparatively small. Taken together, our study reveals the complex set of interrelated conditions that are needed to obtain a breathing pyrochlore distortion in spinel oxides, which accounts for the scarcity of experimental realizations.

\section*{Results and Discussion}

\subsection*{Experimental stability of breathing pyrochlores}
With the goal of expanding the repertoire of spinels that exhibit a breathing pyrochlore lattice, we have attempted to chemically substitute the $B$-site in LiGa$B_4$O$_8$ and LiIn$B_4$O$_8$ with other magnetic transition metals. Specifically, we have attempted to substitute Cr with Mn, Fe, and Co, \textcolor{black}{as outlined in the Methods section.} Despite the compatible ionic radii and valences, none of these substitutions led to the crystallization of the desired breathing pyrochlore structure.  

\begin{figure}[tb]
\includegraphics[width=\columnwidth]{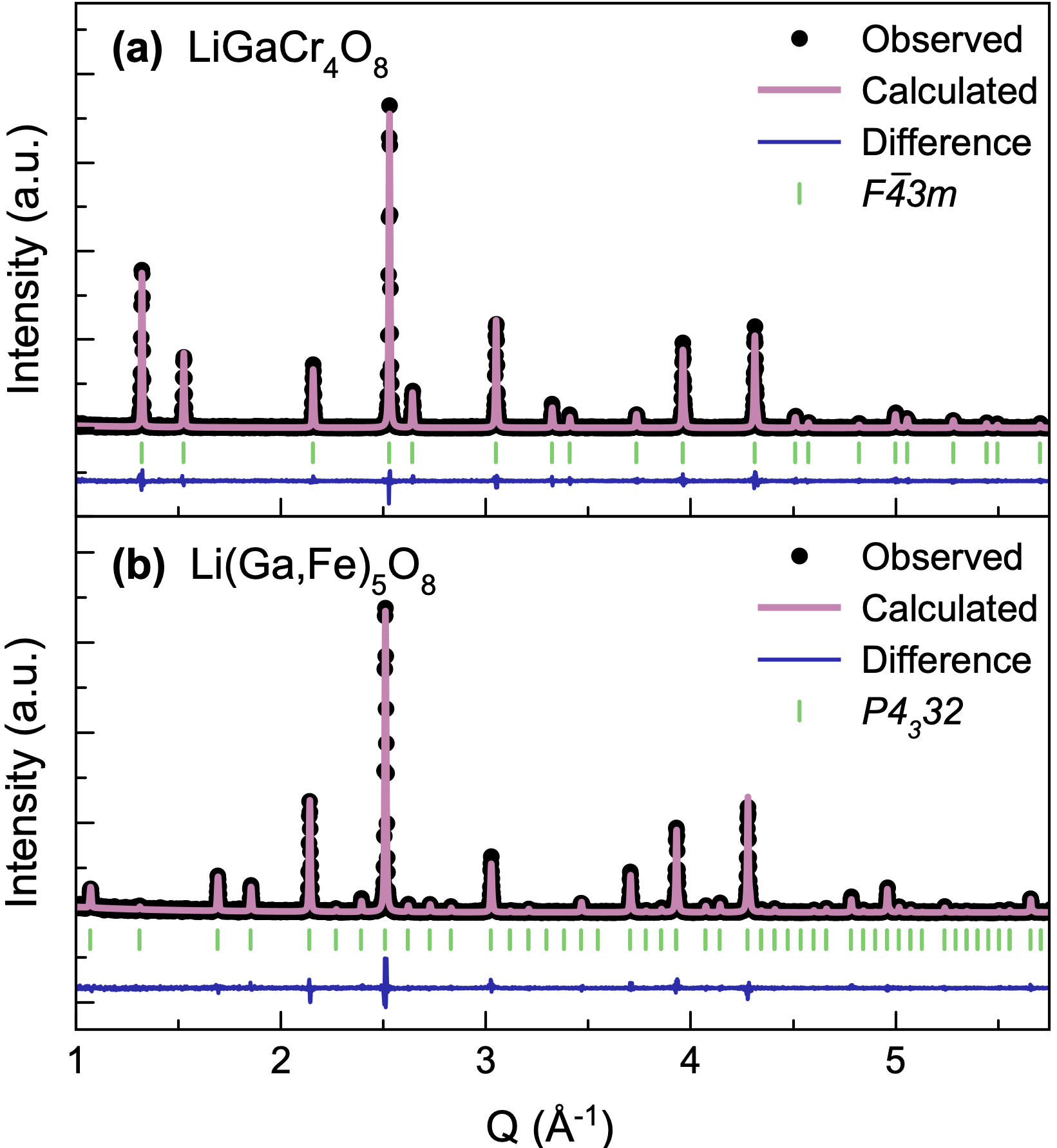}
\caption{\textbf{Successful synthesis of Cr-based breathing pyrochlores and failed synthesis of Fe-based analogs.} Rietveld refinement for the synthesized \textbf{(a)} \(\ch{LiGaCr_4O_8}\) in the cubic $F\overline{4}3m$ breathing pyrochlore space group with goodness of fit parameters $R_{wp} = 26.26$\textcolor{black}{\%}, \textcolor{black}{$R_{exp}= 24.77$\%,} and $GOF = 1.06$ and  \textbf{(b)} \(\ch{Li(Ga_{0.2}Fe_{0.8})_5O_8}\) in the site ordered $P4_332$ cubic spinel derivative structure with goodness of fit parameters $R_{wp} = 25.95$\textcolor{black}{\%}, \textcolor{black}{$R_{exp}= 23.54$\%,} and $GOF = 1.1$. The data are indicated by the black circles, the fit is given by the pink line, and the residual by the blue line. }
 \label{fig:diffraction}
\end{figure}

Before attempting any substitutions, we first
verified the synthesis conditions for the Cr-based breathing pyrochlores. A representative x-ray diffraction (XRD) pattern for LiGaCr$_4$O$_8$ is shown in Fig.~\ref{fig:diffraction}(a). Relative to the diffraction pattern for an undistorted spinel, the $A$-site ordering most clearly manifests through the emergence of superlattice peaks, the first and most intense of which appears at 1.5~\AA$^{-1}$. All observed Bragg peaks can be indexed to the expected $F\overline{4}3m$ space group and no impurity phases were detected. The Rietveld refinement confirms the site ordering of all three cations, as shown in Fig.~\ref{fig:Fig1-SpinelToBreathingPyrochlore}(c). The refined parameters \textcolor{black}{(Appendix Table~\ref{tab:LiGaCr4O8refinement})} are in good agreement with previous reports~\cite{Okamoto2013,saha2017magnetostructural}.

Working our way from left to right across the $3d$ transition metal block, we first discuss the attempted substitution of Cr with Mn. In these reactions, a single-phase crystalline product was not obtained under any reaction conditions and there was no evidence for the formation of a breathing pyrochlore spinel. Instead, a mixture of undistorted spinel (space group $Fd\overline{3}m$) and a phase matching the structure of Mn$_3$O$_4$ (space group $I41/amd$) were observed to form. Mn$_3$O$_4$ is a tetragonally distorted spinel due to the strong Jahn-Teller (JT) distortion of Mn$^{3+}$ in octahedral coordination. While the octahedral oxygen environment surrounding the $B$ cation is minorly distorted in breathing pyrochlore spinels, it is not distorted in a manner that is consistent with the axial elongation preferred by the Mn$^{3+}$ JT effect. We therefore ascribe the JT effect of Mn$^{3+}$ as the primary culprit in the failure of Mn-based breathing pyrochlores to form. 

Moving next to Co, reactions performed in both ambient and oxidizing conditions at a variety of temperatures fail to yield any sign of a breathing pyrochlore phase. We instead observe the formation of common spinels and rock salt phases, indicating that cobalt remains predominantly in the Co$^{2+}$ oxidation state, \textcolor{black}{including when reacted under flowing O$_2$}. It remains possible that more extreme oxidizing conditions, as can be achieved in high-pressure synthesis using oxidizing agents, may favor the formation of a Co-based breathing pyrochlore.

Our final substitution attempt with Fe did result in a single phase crystalline product. This product is not, however, a breathing pyrochlore spinel. We instead detect the formation of a material with the chemical formula \(\ch{Li(Ga_{0.2}Fe_{0.8})_5O_8}\) (space group $P4_332$), as shown in the Rietveld refinement in Fig.~\ref{fig:diffraction}(b). This phase has been previously reported for LiFe$_5$O$_8$\textcolor{black}{, LiGa$_5$O$_8$, and the full solid solution between them}~\cite{Braun1952, Schulkes1963}. This phase is a $B$-site ordered spinel derivative in which the tetrahedral sites are occupied by \textcolor{black}{the trivalent cation} while the octahedral sites are \textcolor{black}{ordered with the remaining trivalent cations and monovalent Li} on two distinct Wyckoff positions.  \textcolor{black}{Interestingly, this space group is both chiral and non-centrosymmetric.} In our \textcolor{black}{Rietveld refinement}, we \textcolor{black}{assume} that Ga is randomly substituted onto both \textcolor{black}{trivalent} sites in proportion to the starting stoichiometry. The full crystallographic details from our refinement are tabulated in the Appendix in Table~\ref{tab:LiGaFe4O8refinement}. We can therefore conclude that in the case of Fe, the breathing pyrochlore phase is unstable towards a different preferred cation ordering.

For our final piece of experimental evidence, we explore the stability of the Cr-based breathing pyrochlores with respect to temperature. Early reports on LiGaCr$_4$O$_8$ claimed that an undistorted $A$-site disordered spinel phase could be obtained by briefly heating the material at 1350~\degree C, well above its synthesis temperature of 1050~\degree C, followed by rapid quenching to room temperature~\cite{Joubert1966,Tarte1973}. We attempted to replicate those findings by conducting experiments across a temperature range of 1200~\degree C to 1500~\degree C, varying the timescales, rate of quenching, and analyzing the resulting XRD patterns. Under none of the conditions we tested, including exact replication of the previous study's method, were we able to isolate a stable, undistorted spinel phase with $A$-site mixing. Instead, at these elevated temperatures, the compound began to show a decrease in the superlattice peaks associated with $A$-site ordering simultaneous with the appearance of peaks corresponding to the corundum-structured precursors \(\ch{Ga_2O_3}\) and \(\ch{Cr_2O_3}\). At the highest temperatures and longest annealing times, the superlattice peaks completely disappear, but never unaccompanied by secondary phases, \textcolor{black}{which in turn alters the stoichiometry of the resulting undistorted spinel}. Less extensive tests were performed on LiInCr$_4$O$_8$ but even with extensive thermal treatments the superlattice peak that originates from $A$- and $A'$-site ordering was not significantly suppressed, even when a partial decomposition into Cr$_2$O$_3$ and In$_2$O$_3$ were observed. These findings suggest that at high temperatures the Cr-based breathing pyrochlores become unstable, leading to decomposition into their precursors but there is a significant difference in the degree of disordering of the $A$-site.

The outcome of our synthesis attempts underscores the fact that the breathing pyrochlore structure, unlike its spinel counterpart, exhibits a notable resistance to substitution, highlighting a gap in our understanding of the structural stability of these materials. Our exploration of the stability of the breathing pyrochlore phase at high temperatures similarly raises questions about the nature of the transition from undistorted spinel to breathing pyrochlore. In the rest of this work, we will take a computational approach to address these questions.

\subsection*{Thermodynamic stability of chemical substitutions in the breathing pyrochlores}

We initiate our computational investigation by first assessing the thermodynamic stability of various substitution attempts on the Cr site using density functional theory (DFT) calculations. \textcolor{black}{The formation energies (enthalpies) reported in this study are computed at 0 K and therefore do not include entropic contributions.} In the case of the Co substitutions, reliable formation energies could not be determined due to the absence of a stable trivalent Co reference oxide. In the case of all other transition metals, the numerical value of the formation energies, presented in Table \ref{table:formation-energies}, can be interpreted to provide further insight into the failed substitution attempts. In the case of Mn, the positive formation energies indicate that these substitutions are not thermodynamically stable. As previously mentioned, this instability can be understood as originating from the strong Jahn-Teller effect, which makes an undistorted octahedral environment unfavorable for Mn$^{3+}$. Interestingly, the formation energies of the Cr and Fe-based breathing pyrochlores are similar, indicating that all four compounds have the potential to be thermodynamically stable. This suggests that thermodynamic stability is not a limiting factor for the formation of Fe-based breathing pyrochlores, highlighting the fact that the failed substitutions involving Fe must be attributed to factors beyond simple thermodynamic considerations.

\begin{table}[tb]
    \caption{Formation energies per formula unit (F.U.) of various substitution attempts of the Cr ion at the \(B\)-site in the breathing pyrochlores LiGa$B_4$O$_8$ and LiIn$B_4$O$_8$ ($B=$~Cr, Mn, Fe, and Co). The formation energies of Co-based compounds were omitted due to the variable oxidation state of Co, which affected the reliability of the DFT calculations. }
    \centering
    \begin{tabular}{l>{\centering\arraybackslash}p{3cm}>{\centering\arraybackslash}p{2.7cm}}
        \toprule
        Material & \makecell{Experimentally \\ observed (y/n)} & \makecell{Formation \\ Energy (eV/F.U.)}\\
        \hline
        \ch{LiGaCr_4O_8} & y &  -0.980 \\
        \ch{LiInCr_4O_8} & y & -0.668 \\
        \ch{LiGaMn_4O_8} & n & 0.818 \\
        \ch{LiInMn_4O_8} & n & 0.884 \\
        \ch{LiGaFe_4O_8} & n & -0.959 \\
        \ch{LiInFe_4O_8} & n & -0.762 \\
        \ch{LiGaCo_4O_8} & n & -- \\
        \ch{LiInCo_4O_8} & n & -- \\
        \toprule
    \end{tabular}
    \label{table:formation-energies}
\end{table}

\begin{figure}[tbp]
\includegraphics[width=1\columnwidth]{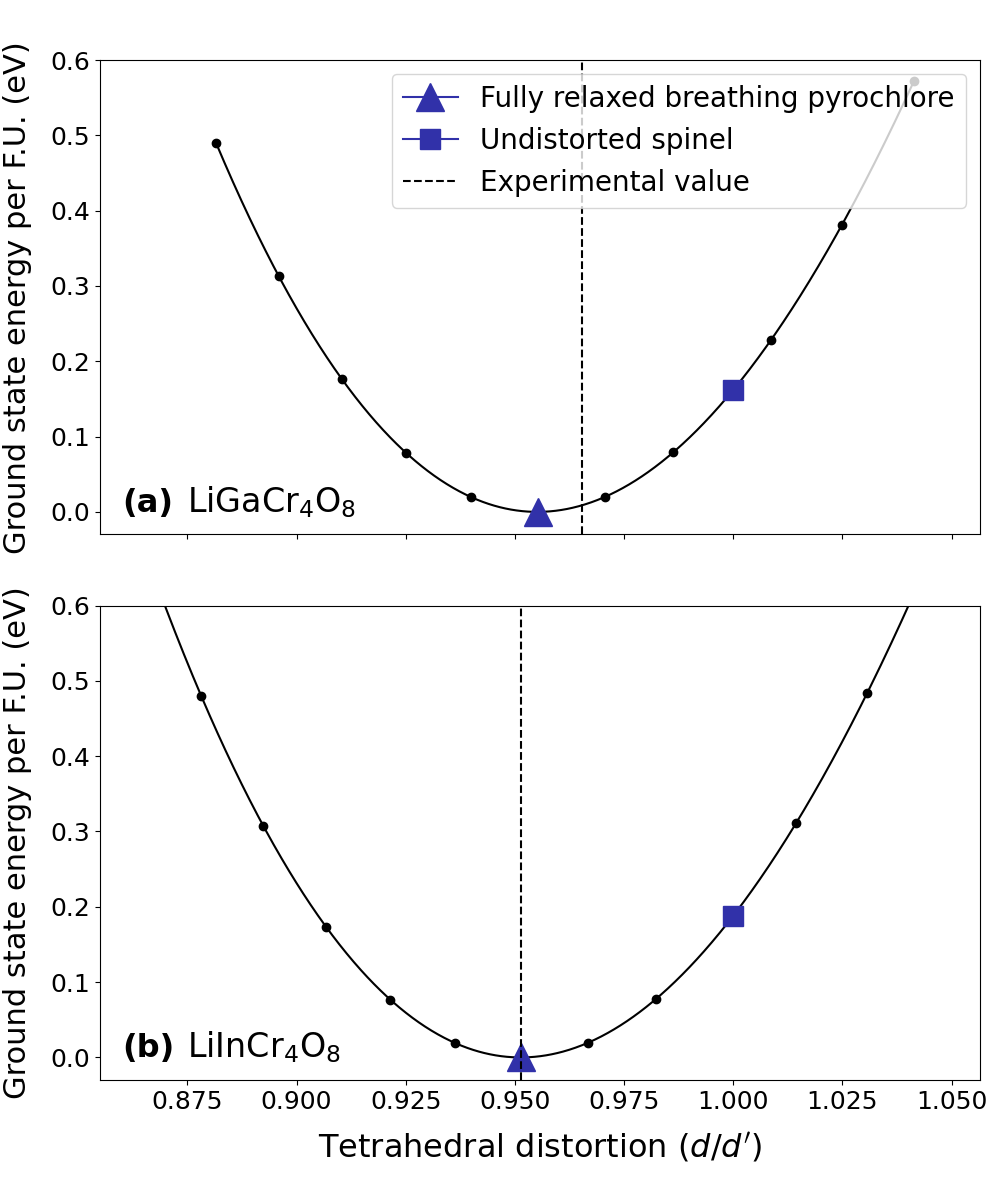}
\caption{\textcolor{black}{\textbf{Energetics of the tetrahedral distortion in site ordered spinels.} The ground state energy per formula unit (F.U.) for \textbf{(a)} LiGaCr$_4$O$_8$ and \textbf{(b)} LiInCr$_4$O$_8$ with varying tetrahedral distortion ($d/d'$) starting from the $A$ and $A'$ site ordered conditions, which reach a minimum at $d/d' = 0.955$ and $d/d' = 0.951$, respectively, corresponding exactly to the tetrahedral distortions of the relaxed structures. The vertical dashed lines indicate the experimentally determined tetrahedral distortions. We also note the parabolic relationship between energy and tetrahedral distortion, meaning that there is no energetic barrier for this displacive transition.}}
\label{fig:SelectiveDynamics-BreathingPyrochlore-Distortion}
\end{figure}

An interesting question arises when considering how the energetics of breathing pyrochlores compare to those of an undistorted spinel structure that retains $A$-site order but lacks the characteristic breathing distortion (as pictured in Fig.~\ref{fig:Fig1-SpinelToBreathingPyrochlore}(b)). 
\textcolor{black}{To explore this question and estimate the energy scale of the tetrahedral distortion, we performed DFT calculations using the selective dynamics modality. Within this approach, all atoms in the system were allowed to relax except for the Cr ions forming the breathing pyrochlore lattice, which were fixed at prescribed positions corresponding to varying tetrahedral distortions. By systematically varying the tetrahedral distortion while preserving the overall symmetry of the system we established a clear relationship between the ground-state energy and tetrahedral distortion, as shown in Figure \ref{fig:SelectiveDynamics-BreathingPyrochlore-Distortion}. For both LiGaCr$_4$O$_8$ and LiInCr$_4$O$_8$, the energy minimum aligns well with the experimentally observed tetrahedral distortion.
For LiGaCr$_4$O$_8$, the energy difference between the undistorted site-ordered spinel structure ($d/d' = 1$) and the experimentally observed breathing pyrochlore structure ($d/d' = 0.955$) is \( 0.162 \, \text{eV}\) per F.U. 
Similarly, for LiInCr$_4$O$_8$, we observe an energy difference of \(0.189 \, \text{eV}\) per F.U between the breathing pyrochlore and the undistorted spinel structure retaining $A$/$A'$-site order. }

\subsection*{Breathing pyrochlore $A$-site energetics}

The initial focus of our investigation is to understand the relationship between $A$-site ordering and the nature of the distortion of the pyrochlore sublattice in \(\ch{LiGaCr_4O_8}\) and \(\ch{LiInCr_4O_8}\) breathing pyrochlores. To this end, we first assess the energy scale associated with the charge ordering of the $A$ and $A'$ cations. We conducted a comparative analysis of the ground-state energy for two distinct structures: first, the conventional unit cell of \(\ch{LiGaCr_4O_8}\) and \(\ch{LiInCr_4O_8}\), with Cr atoms exclusively occupying the octahedral $B$ sites and Li and Ga or In ions exclusively occupying the tetrahedral $A$ and $A'$ sites, respectively. Second, we examined a modified structure where a single Li atom is swapped with a Ga or In atom, resulting in a Li atom occupying an $A'$ site and a Ga or In atom occupying an $A$ site. By analyzing the ground state energy difference between these two configurations, we assess the energy scale associated with $A$-site ordering. 

\begin{figure}[tbp]
\includegraphics[width=\columnwidth]{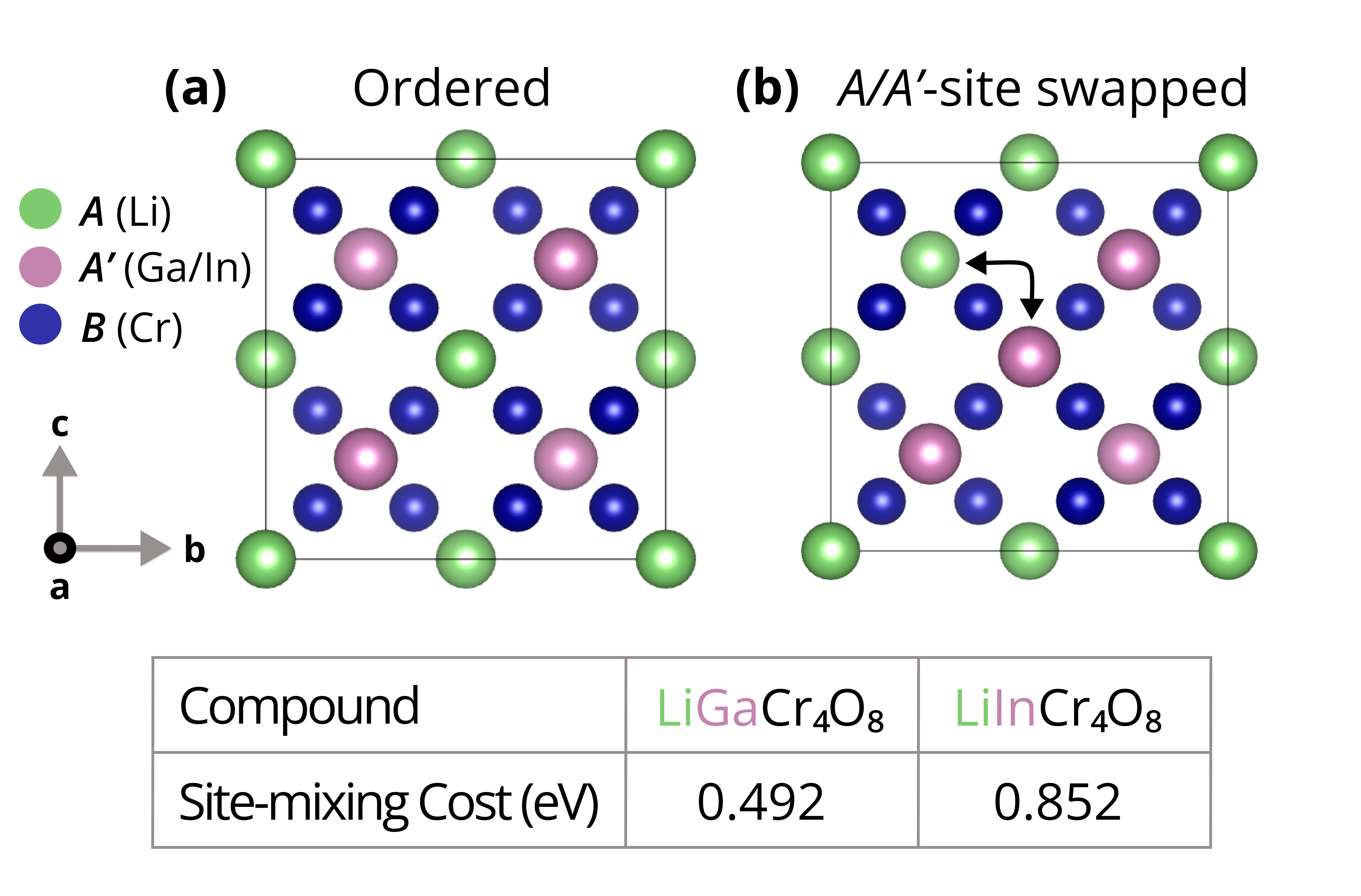}
\caption{\textbf{Energetics associated with $A$-site order in the breathing pyrochlores.} Comparison of \textbf{(a)} the pristine pyrochlore lattice in \(\ch{LiGaCr_4O_8}\) and \(\ch{LiInCr_4O_8}\) with \textbf{(b)} a disordered breathing pyrochlore structure for which a single \(A/A'\)-site swap was performed. In the ordered structure, Li$^+$ (green) occupies the \(A\)-site, while Ga$^{3+}$ or In$^{3+}$ (pink) occupies the \(A'\) site and Cr$^{3+}$ (blue) occupies the \(B\) site. In the disordered structure,  one \ch{Li} is swapped with \ch{Ga} (or \ch{In}) on the \(A\) and \(A'\) sites, leading to a mixed occupancy. The swapping cost, represented by the energy difference between the ground state energy of the two structures, quantifies the energetic penalty for this site exchange.}
 \label{fig:Fig2-A-Site-Mixing}
\end{figure}

The analysis, visually represented in Figure \ref{fig:Fig2-A-Site-Mixing}, reveals that the energy difference between the two aforementioned structures, and therefore the cost of displacing one $A$-site cation to the $A'$-site and vice versa, is 0.492 eV and 0.852 eV for \(\ch{LiGaCr_4O_8}\) and \(\ch{LiInCr_4O_8}\), respectively.
The numerical value of these site-mixing costs becomes insightful when considering the Boltzmann statistics at the synthesis temperature, where the probability of an $A$-site being occupied by an $A'$-site ion (and vice versa) at a temperature \(T\) is proportional to $e^{\sfrac{-\Delta E}{k_B T}}$, with \(\Delta E\) as the ground state energy difference between the two structures, and \(k_B\) as the Boltzmann constant. Given the substantial $A/A'$-site mixing cost for \(\ch{LiInCr_4O_8}\), the probability of site mixing is exceedingly small: at 1350~\degree C, a temperature at which experimentally the material begins to decompose, the site mixing probability is just 0.2\%. In the case of LiGaCr$_4$O$_8$, the smaller site-mixing cost corresponds to a higher probability of 1\% at the synthesis temperature of 1050~\degree C, which grows to 3\% at the decomposition temperature of 1350~\degree C. It is interesting to reemphasize here that experimentally heat treatments at 1350~\degree C completely suppress the breathing pyrochlore superlattice peak in the case of LiGaCr$_4$O$_8$ but not in the case of LiInCr$_4$O$_8$, which appears consistent with the analysis shown here.

We now consider the nature of the breathing pyrochlore distortion in $A$-site-ordered spinels. \textcolor{black}{The energy scale associated with perturbing the $A$-site cation order is significantly larger than that associated with the tetrahedral distortion, as shown in Fig.~\ref{fig:SelectiveDynamics-BreathingPyrochlore-Distortion}. Once the $A$-site cation order is in place, we see that there is no energetic barrier toward the displacive distortion of the Cr-sublattice.} Moreover, the large energetic barrier to site mixing between the $A$- and $A'$-sites supports the notion that $A$-site ordering is present at all experimentally relevant temperatures and drives the breathing pyrochlore distortion. In energetic terms, this transformation can be visualized as a three-step process: starting from a spinel with an undistorted pyrochlore lattice and disordered $A$ and $A'$ cations as shown in Fig.~\ref{fig:Fig1-SpinelToBreathingPyrochlore}(a), upon A-site ordering as shown in Fig.~\ref{fig:Fig1-SpinelToBreathingPyrochlore}(b), a necessary condition is met for the subsequent distortion of the pyrochlore lattice to occur, resulting in the breathing pyrochlore structure in Fig.~\ref{fig:Fig1-SpinelToBreathingPyrochlore}(c). This scenario is consistent with the picture presented by Ref.~\cite{Okamoto2013}, wherein the $A$ site ordering was explained as resulting from minimization of electrostatic energy arising from the difference in the valence states between \(\ch{Li^+}\) and \(\ch{Ga^{3+}}\)/\(\ch{In^{3+}}\).
 
\subsection*{Breathing pyrochlore $B$-site energetics}

\begin{figure*}
\centering
\includegraphics[width=\textwidth]{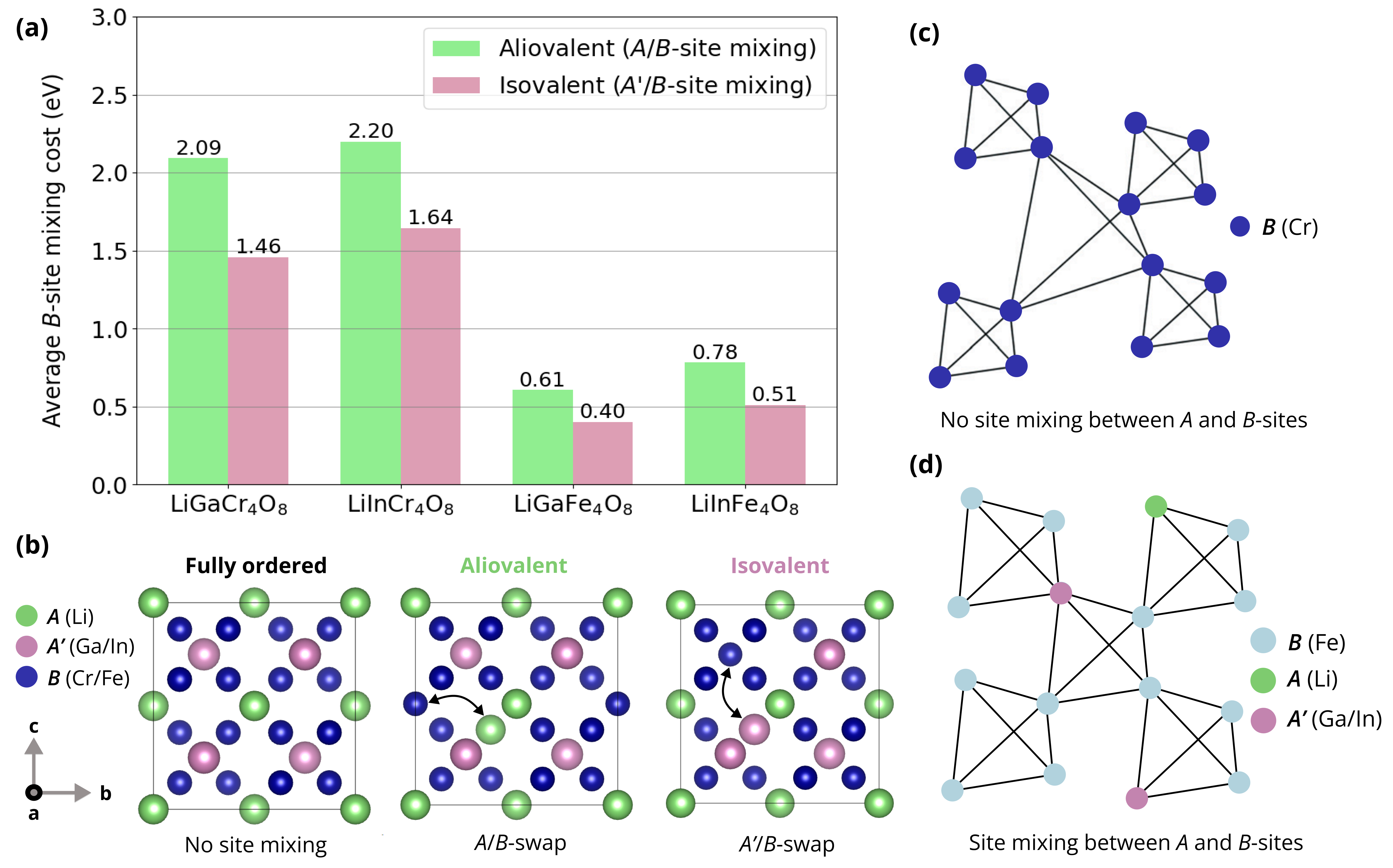}
 \caption{\textbf{Energetics associated with $B$-site mixing in the breathing pyrochlores.} \textbf{(a)} The energetic cost to site mixing between the $B$-site cation (Cr$^{3+}$ or Fe$^{3+}$) with the $A$ or $A'$ site cation. The costs are categorized into isovalent (In$^{3+}$/Ga$^{3+}$) and aliovalent (Li$^{+}$) swaps. For all compositions, we observe a larger cost to making an aliovalent swap as compared to the isovalent case, highlighting the importance of charge order in these materials. We can also observe the approximately five-times smaller cost of performing these swaps in the case of Fe as compared to Cr due to the differences in their octahedral site preference energies. \textbf{(b)} Schematics of the different site mixing scenarios: fully ordered with no site mixing, aliovalent swap (Cr-Li), and isovalent swap (Cr-Ga/In). As a consequence of the energetics presented here, the integrity of the ordered sublattices is maintained in the case of \textbf{(c)} the Cr-based samples allowing the breathing distortion to occur while in the case of \textbf{(d)} the Fe based materials, significant site mixing is permitted and no breathing distortion is observed.}
\label{fig:Fig3-BSite-Mixing}
\end{figure*}

The analysis in the previous section suggests that the energy scale of the $A$-site ordering likely exceeds the energy scale of the breathing distortion itself in LiGaCr$_4$O$_8$ and LiInCr$_4$O$_8$.
However, this observation does not distinguish between Cr and Fe, thus failing to clarify why substitution attempts at the \(B\)-site did not yield stable structures. To further investigate this, we examine additional factors associated with the $B$ sublattice, beginning with the octahedral site preference energy (OSPE) of the $3d$ transition metal ions in the breathing pyrochlore framework. The stability of \(\ch{Cr^{3+}}\) in an octahedral environment is notably high due to its half-filled \(t_{2g}\) orbitals, resulting in a strong preference for octahedral sites over tetrahedral sites. In contrast, \ch{Fe^{3+}}, with its \(d^5\) electronic configuration, shows no significant preference for either octahedral or tetrahedral sites making it an ideal model system for understanding the effect of the OSPE of the $B$-site ion on the overall energetics of breathing pyrochlore spinels. 

The OSPE was obtained by calculating the crystal field splitting \(\Delta_{o}\) using a tight-binding model constructed with maximally localized Wannier functions (MLWFs) centered on the \(d\)-orbitals of \ch{Cr^{3+}} in the primitive structure of LiGaCr$_4$O$_8$. As the crystal field splitting \( \Delta_{o} \) varies minimally among metals with similar ionic valencies \cite{Dunitz1967}, and is expected to remain relatively constant for $3d$ transition metals in the same geometry,  we extend the value of $\Delta_{o}$ from \ch{LiGaCr_4O_8} in all subsequent discussions regarding the OSPE. On-site energies for each of the five $d$ orbitals are tabulated in Table~\ref{table:onsite-energies} in the Appendix, and the resulting crystal field splitting \(\Delta_{o}\) was determined to be \(2.2476  \pm 0.0001 \,\text{eV}\). From this, the OSPE for \(\ch{Cr^{3+}}\) was subsequently calculated to be -\(1.8980 \pm 0.0001\, \text{eV}\). In contrast, the OSPE for \(\ch{Fe^{3+}}\) is zero due to its \(d^5\) electronic configuration, indicating no preference for octahedral or tetrahedral sites.

Taking the OSPE values for \(\ch{Cr^{3+}}\) and \(\ch{Fe^{3+}}\), we examined their implications on the overall energy of the breathing pyrochlore system through a site-mixing analysis. We compared the ground-state energies of three distinct crystal configurations shown in Fig.~\ref{fig:Fig3-BSite-Mixing}(b): (i) a standard breathing pyrochlore unit cell, (ii) a structure where a single Cr or Fe cation is swapped with Li, and (iii) a structure where a Cr or Fe cation is swapped with Ga or In. In both swapped configurations, one Cr or Fe occupies a tetrahedral site instead of an octahedral one but iii) is isovalent whereas ii) is aliovalent. Rather than limiting our analysis to a single swapped configuration, we expand our investigation to include multiple configurations. This approach allows for a robust evaluation of site-mixing effects on the breathing pyrochlore's stability, capturing potential systematic uncertainties that may affect the analysis. Specifically, we fix a \(\ch{Cr^{3+}}\) (or \(\ch{Fe^{3+}}\)) ion within the conventional unit cell and generate four configurations (some of them symmetry equivalent) by swapping it with each of the four Li ions. This procedure is repeated for Ga and In, ensuring thorough coverage of both \(A\)/\(B\)-site swaps and \(A'\)/\(B\)-site swap scenarios. The resulting average $B$-site mixing cost for all permutations are presented in Fig.~\ref{fig:Fig3-BSite-Mixing}(a).

From examination of the site mixing energies in Fig.~\ref{fig:Fig3-BSite-Mixing}(a) a few trends are immediately apparent. First, in all cases the isovalent swap has a lower energy cost that the corresponding aliovalent swap. This can be understood in the context of the $A$/$A'$ cation ordering described in the previous section, where the isovalent swap preserves the underlying charge order. In the case of the Cr-based materials we can see that the difference between the isovalent ($A'$/$B$) and aliovalent ($A$/$B$) swaps correlates very well with the observed ($A$/$A'$) costs presented in Fig.~\ref{fig:Fig2-A-Site-Mixing}. The second, and most striking feature, is that the site mixing cost for the Fe-based compositions is substantially lower in all cases than the Cr-based compositions. The close alignment of the average site-mixing costs in LiGaCr$_4$O$_8$ and LiInCr$_4$O$_8$  with the \textit{ab-initio} computed 
value of the OSPE for \(\ch{Cr^{3+}}\) (-\(1.8980 \pm 0.0001\, \text{eV}\)) suggests that the octahedral site preference of Cr primarily drives the energy cost associated with site mixing. In contrast, the markedly lower site-mixing cost in Fe-based compositions suggests that cation disorder, particularly between the \(A'\) and \(B\)-sites is much more favorable in these compounds due to the zero OSPE of Fe. The nonzero cost of mixing in the isovalent case for the Fe-based compositions can be understood as originating from size effects, where the lower site mixing cost in LiGaFe$_4$O$_8$ is due to the more similar ionic radii of Fe$^{3+}$ and Ga$^{3+}$ as compared to the larger mismatch between Fe$^{3+}$ and In$^{3+}$.

To further understand the thermodynamic likelihood of \(B\)-site ordering in Cr versus Fe-based breathing pyrochlores, we again performed a Boltzmann statistical analysis. Here, we consider the energy cost of the isovalent swap, an approach that allows us to preserve electrostatic considerations while focusing solely on the thermodynamic aspects. At the synthesis temperature of 1050 \degree C, the probability of site-disorder between the \(B\)-sites and the \(A'\) (Ga/In)-sites for Cr-based breathing pyrochlores is extremely low, of order 0.0001\%. In contrast, for Fe-based breathing pyrochlores, Boltzmann statistics suggest a significantly higher probability of \(B\)-site disorder at the synthesis temperature, with probabilities of order 1\% corresponding to four orders of magnitude increase in likelihood over their Cr-analogs.

The differing values of the site-mixing costs for Cr and Fe-based breathing pyrochlores, presented in Fig.~\ref{fig:Fig3-BSite-Mixing}(a), become very insightful when considered in relation to the formation of the breathing pyrochlore lattice. Due to the very high energetic barrier to site mixing, the pyrochlore lattice depicted in Fig.~\ref{fig:Fig3-BSite-Mixing}(c) will only contain Cr atoms at the \(B\)-site, thus preserving its structural integrity. In contrast, in the case of Fe, there is a significantly reduced energy penalty associated with displacing Fe from its octahedral site (as suggested by the above Boltzmann analysis), leading to site-mixing in Fe-based spinels being much more probable. Consequently, the pyrochlore lattice displayed in Fig.~\ref{fig:Fig3-BSite-Mixing}(d) will feature a mix of Fe, Li, and Ga (or In) cations across both the $A$ and $B$ sublattices. Such mixing can disrupt the integrity of the pyrochlore lattice, disrupting the $A/A'$ and $B$-site order and thereby inhibiting the characteristic lattice distortion essential for its formation.

\subsection*{Magnetic exchange interactions in the breathing pyrochlores}

\begin{figure}[tbp]
  \centering
      \includegraphics[width=\columnwidth]{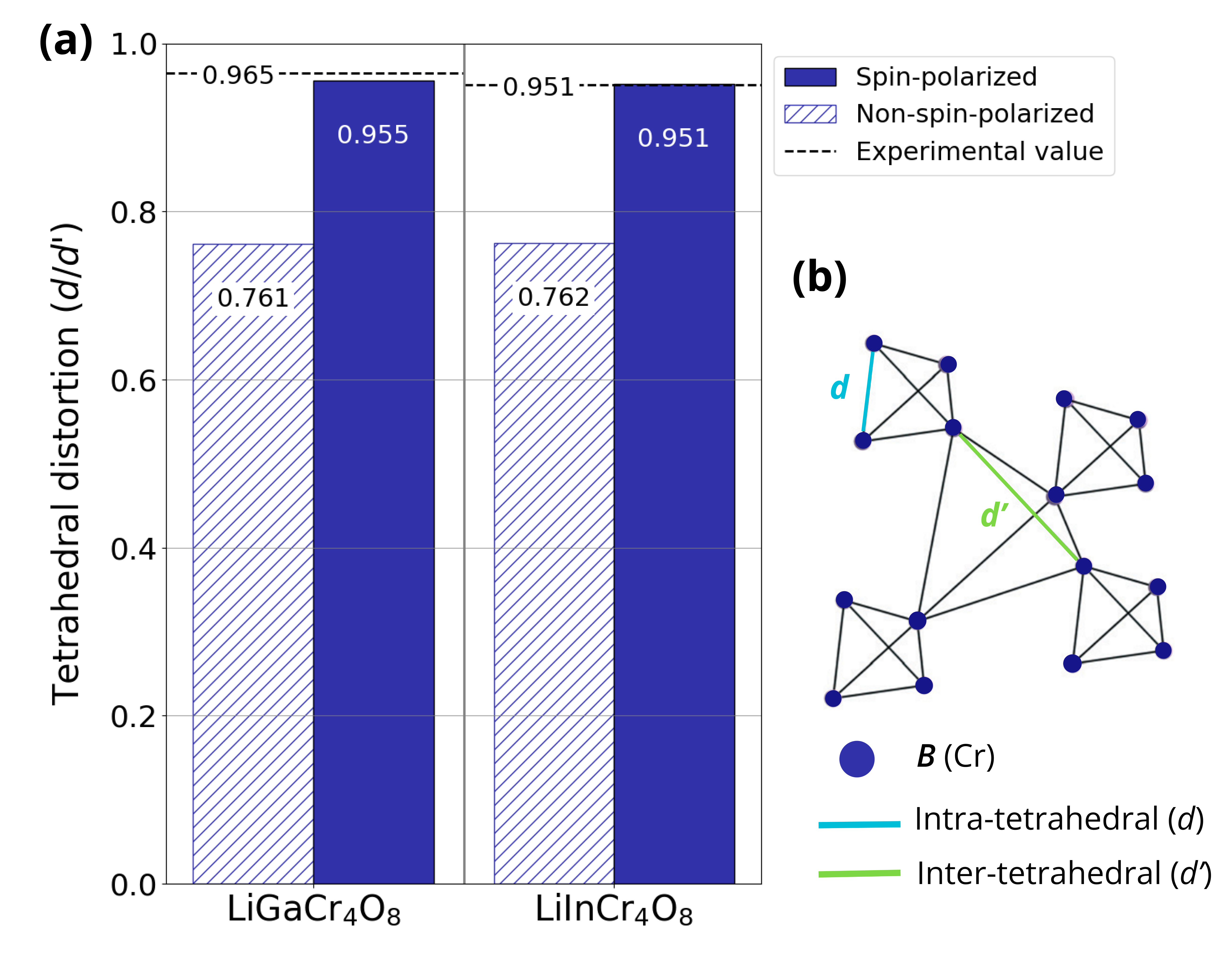}
   \caption{ \textbf{The magnitude of the \textcolor{black}{tetrahedral} distortion can only be accurately captured using spin-polarized DFT.} \textbf{(a)} Tetrahedral distortion (defined as $\sfrac{d}{d'}$, where $d$ and $d'$ are the intra- and inter-tetrahedral distance, respectively, as shown in \textbf{(b)}) for the relaxed structures of \ch{LiGaCr_4O_8} and \ch{LiInCr_4O_8} under spin-polarized and non-spin-polarized DFT calculations. The bars represent the calculated breathing ratios for the relaxed structures, with solid bars indicating spin-polarized DFT calculations and hatched bars indicating non-spin-polarized DFT calculations. The horizontal dashed lines represent the experimental tetrahedral distortion for comparison from Ref.~\cite{Okamoto2013}.}
\label{fig:Fig4-Distortion}
\end{figure}

First-principle calculations were carried out to elucidate the role of magnetism in the formation of the \(\ch{LiGaCr_4O_8}\) and \(\ch{LiInCr_4O_8}\) breathing pyrochlores. Our initial assessment focused on evaluating the effects of non-spin-polarized and spin-polarized DFT calculations on these materials. We find that the ground state spin arrangement for both \(\ch{LiGaCr_4O_8}\) and \(\ch{LiInCr_4O_8}\) is antiferromagnetic (AFM), consistent with previous experimental and computational studies \cite{Okamoto2013,Ghosh2019}. 
Significantly, our results reveal that the relaxed structure accurately reproduces the experimentally determined structure only when spin-polarized DFT calculations are employed. The most striking example of this is the magnitude of the tetrahedral distortion, $\sfrac{d}{d'}$, as shown in Figure \ref{fig:Fig4-Distortion}. In the case of the non-spin-polarized calculations, the magnitude of the distortion is dramatically overestimated (larger deviation from an undistorted pyrochlore lattice with $\sfrac{d}{d'} = 1$). In contrast, the spin-polarized calculations show striking agreement with the experimental value. Similar results are obtained for the Cr-O-Cr angle (which we will denote as \(\alpha\)). The numerical values for \(\sfrac{d}{d'}\) and \(\alpha\) for both spin-polarized and non-spin-polarized DFT, alongside the experimental values from crystallographic databases, are provided in Table~\ref{table:supplementaryInformation-breathingpyrochlore-distortion} of the Appendix. Importantly, the type of spin arrangement (FM or AFM) does not significantly affect the tetrahedral distortion. These findings suggest that the spin degree of freedom plays a crucial role in determining the magnitude of the tetrahedral distortion for the breathing pyrochlores \ch{LiGaCr_4O_8} and \ch{LiInCr_4O_8}.

Next, we extract the isotropic nearest-neighbor magnetic exchange interactions from the spin Heisenberg model for \ch{LiGaCr_4O_8} and \ch{LiInCr_4O_8}. In the case of breathing pyrochlores, which feature a three-dimensional lattice of corner-sharing tetrahedra of alternating size, the spin Hamiltonian is defined as
\begin{equation}
  H = J \sum_{\langle ij \rangle} S_i S_j + J' \sum_{\langle ij \rangle} S_i S_j,
\label{eq:BreathingPyroSpinHamiltonian}
\end{equation}
where \(J\) represents the nearest-neighbour interactions within the small tetrahedra, and \(J'\) represents those within the large tetrahedra. The summation over \( ij\) in both the first and second term of Equation \ref{eq:BreathingPyroSpinHamiltonian} runs over the Cr-Cr bonds within the small and large tetrahedra, respectively. The magnetic exchange interactions were extracted using the four-state method \cite{Sabani2020, Xiang2011, Xiang2013}. The magnitude and ratio of the magnetic exchange interactions \(J\) and \(J'\), obtained via the energy mapping method, vary significantly depending on the effective Hubbard \(U\) term applied to the Cr \(d\)-orbitals (see Appendix Fig.~\ref{fig:Ueff_magnetic_exchange_breathing_factor}). Accordingly, we selected a single Hubbard correction of $U=1$~eV to align with the experimentally determined breathing ratio for both \(\ch{LiInCr_4O_8}\) and \(\ch{LiGaCr_4O_8}\). The resulting exchange couplings are presented in Table \ref{table:exchange-constants}, along with the experimental~\cite{Okamoto2013,he2021neutron} and computational values~\cite{Ghosh2019}.

Our computed exchange coefficients are in reasonable agreement with the experimental values in both materials. However, it is worth emphasizing that our spin-polarized DFT study was confined to collinear magnetic configurations, which might not fully capture the complex magnetic ground states of \ch{LiGaCr_4O_8} and \ch{LiInCr_4O_8}, which also involve structural degrees of freedom~\cite{nilsen2015complex,Lee2016}. A notable difference between our findings and the previously computed values~\cite{Ghosh2019} is in the relative size and of $J$ and $J'$ for LiGaCr$_4$O$_8$. In this case, methodological differences, particularly the further neighbor couplings included in the Hamiltonian of Ghosh \emph{et al.}, likely contribute to the observed discrepancies. As our primary focus is on the overall magnitude of magnetic exchange interactions in relation to the breathing pyrochlore structural distortion, we do not investigate these differences further. Most importantly, our results indicate that the magnetic exchange interactions are on the meV energy scale, orders of magnitude lower than the energy scales associated with \(A\)-site and \(B\)-site ordering. Consequently, these magnetic exchange interactions are unlikely to significantly influence the structural stability of breathing pyrochlores. However, the presence of magnetic degrees of freedom may play a significant role in setting the magnitude of the breathing pyrochlore distortion.

\begin{table}[tbp]
\caption{Computed exchange constants \(J\) and \(J'\) (in meV) and breathing factors \(B_f=\sfrac{J'}{J}\) for the breathing pyrochlores \ch{LiGaCr_4O_8} and \ch{LiInCr_4O_8} along with previous experimental and computational reports.}
\centering
\begin{tabular*}{\columnwidth}{@{\extracolsep{\stretch{1}}}*{4}{c}@{}}
\toprule
\multicolumn{4}{c}{\textbf{\ch{LiGaCr_4O_8}}} \\
\hline
\textbf{} & \bm{$J$} \textbf{(meV)} & \bm{$J'$} \textbf{(meV)} & \textbf{$B_f=\sfrac{J'}{J}$} \\
\hline
Computed (this work) & 7.74 & 5.30 & 0.68 \\
Computed \cite{Ghosh2019} & 5.70 & 8.62 & 1.51 \\
Experimental \cite{Okamoto2013}& 5.18 & 2.59 & 0.50 \\
Experimental \cite{he2021neutron}& 10.4 & 6.2 & 0.60 \\
\hline
\multicolumn{4}{c}{\textbf{\ch{LiInCr_4O_8}}} \\
\hline
\textbf{} & \bm{$J$} \textbf{(meV)} & \bm{$J'$} \textbf{(meV)} & \textbf{$B_f=\sfrac{J'}{J}$} \\
\hline 
Computed (this work)& 7.01 & 0.90 & 0.13 \\
Computed \cite{Ghosh2019} & 5.16 & 1.90 & 0.37 \\
Experimental \cite{Okamoto2013} & 4.31 & 0.52 & 0.12 \\
\toprule
\end{tabular*}
\label{table:exchange-constants}
\end{table}

\section*{Summary and Outlook}

In this paper, we have investigated the energy scales of three distinct phenomena in breathing pyrochlore spinels of the form $AA'B_4$O$_8$: the energy scale associated with $A$-site charge ordering, the energy scale related to $A$/$B$-site mixing, and the energy scale of the magnetic exchange interactions.
In particular, we explain why \(\ch{LiGaCr_4O_8}\) and \(\ch{LiInCr_4O_8}\) breathing pyrochlores exist, while the Fe-based analogs do not. The critical precondition for forming a breathing pyrochlore lattice is the strong octahedral site preference of the transition metal, which is maximized in the case of Cr and zero in the case of Fe. As a result, near-perfect ordering of the pyrochlore sublattice occurs for Cr while significant $A$/$B$-site mixing is allowed with Fe. Once the precondition of $B$-site order is met, we find that the charge ordering of the $A$ and $A'$ cations is energetically preferred at all temperatures up to the decomposition temperature. Consequently, significant $A$/$A'$ site mixing is never expected under any realistic experimental conditions in LiGaCr$_4$O$_8$ and LiInCr$_4$O$_8$. While the magnetic exchange interactions between the Cr$^{3+}$ cations are significantly overshadowed by the other energy scales discussed here, we nonetheless find that they play a deterministic role in setting the magnitude of the breathing pyrochlore distortion.

The stringent requirement on the OSPE of the transition metal cation naturally accounts for the scarcity of spinels with breathing pyrochlore distortions. The simultaneous constraints of a strong octahedral site preference energy and a trivalent oxidation state provide relatively few options. Within the $3d$ transition metal block, the maximal OSPE of Cr$^{3+}$ with a $d^3$ electron configuration can only be matched by a $d^8$ electron configuration, which in a trivalent state is achieved only for Ni$^{3+}$. This particular oxidation state of Ni has recently been explored in NdNiO$_3$ as the precursor to thin film nickelate superconductors~\cite{li2019superconductivity} but is infamously difficult to stabilize in bulk materials~\cite{zhang2017high}. Synthesis under extreme conditions may open a path to stabilizing a nickelate breathing pyrochlore. Very large OSPEs are also achievable for cations with low spin states, which among oxides are usually only found for $4d$ and $5d$ transition metals. This exact situation is realized in the rhodium breathing pyrochlores~\cite{Joubert1966} but the resulting materials are non-magnetic due to the filled $t_{2g}$ state of Rh$^{3+}$. A comparably strong OSPE is obtained for Ru$^{3+}$ with a low-spin $d^5$ electron configuration, making this a particularly attractive target.

While our investigation here has been limited to spinel oxides, these findings can naturally be extended to the analogous sulfide and selenide systems, $AA'B_4$S$_8$ and $AA'B_4$Se$_8$. Due to the more spatially extended $3p$ and $4p$ orbitals of these anions, the increased covalency results in a stronger ligand field splitting of the transition metal $d$ states. Therefore, the role of the octahedral site preferences will be further exacerbated. Accordingly, in the selenides and sulfides, Cr is again the only magnetic transition metal known to fully inhabit the $B$ sublattice. However, in this set of materials, the enlarged lattice parameters permit a larger variety of $A$ and $A'$ cations, including $Cu^+$, Ag$^+$, and $Al^{3+}$~\cite{pinch1970some}, providing a pathway to different breathing ratios, exchange interactions, and resulting magnetic ground states~\cite{Ghosh2019,reig2021frustrated}. One striking feature of the sulfide and selenide materials is that in a couple of instances, specifically AgAlCr$_4$S$_8$ and AgGaCr$_4$S$_8$, charge ordering onto $A$ and $A'$ sites is avoided. This may indicate that with increasing metallicity the electrostatic constraints that enforce charge order are weakened.

\section*{Methods}
\subsection*{Experimental}
Polycrystalline samples of the breathing pyrochlore \(\ch{LiGaCr_4O_8}\) and \(\ch{LiInCr_4O_8}\)were prepared by solid state reaction. \(\ch{Li_2CO_3}\), \(\ch{Ga_2O_3}\) (or In$_2$O$_3$), and \(\ch{Cr_2O_3}\) were mixed with 10\% excess of \(\ch{Li_2CO_3}\), pressed into pellets, and fired at 1000~\degree C and 1050~\degree C for 24 h each, \textcolor{black}{followed by slow cooling to room temperature,} with intermediate regrinding and repressing. Attempts more closely following Okamoto \emph{et al.} (no excess \(\ch{Li_2CO_3}\) with two sinterings at 1000~\degree C and 1100~\degree C for 24 h~\cite{Okamoto2013}) resulted in a sample with corundum-structured \(\ch{Ga_2O_3}\) and/or \(\ch{Cr_2O_3}\) impurities. Various substitutions (replacement of Cr with Fe, Co, or Mn) were attempted following a similar procedure. In the specific case of Fe, stoichiometric ratios of \(\ch{Li_2CO_3}\), \(\ch{Ga_2O_3}\) and \(\ch{Fe_2O_3}\) were mixed and pelleted, followed by two sinterings at 1000~\degree C and 1100~\degree C for 24 hours each with intermediate regrinding and repressing.

In an attempt to isolate an $A$-site disordered spinel, both \(\ch{LiGaCr_4O_8}\) and \(\ch{LiInCr_4O_8}\) were subjected to high temperature treatments. The phase pure samples were heated to temperatures in the range of 1200~\degree C to 1500~\degree C, with varying timescales from 5 min to 20 h. To accurately simulate the conditions of previous studies~\cite{Joubert1966,Tarte1973}, the \(\ch{LiGaCr_4O_8}\) samples were inserted at the already preheated furnace and then quenched, either in air or by dropping the pellet onto a metallic tray to improve heat dissipation.

The phase composition of the synthesized materials was confirmed using powder x-ray diffraction. Measurements were performed on a Bruker D8 Advance diffractometer using a Cu x-ray source with monochromated $\lambda_{K\alpha_1} = 1.5418$~\AA~in Bragg-Brentano geometry at room temperature. Rietveld refinements were performed using TOPAS~\cite{coelho2018topas}. Refined quantities include the background, sample displacement, lattice constants, atomic positions, site occupancies, modified peak shape, strain broadening, and thermal parameters.

\subsection*{Density functional theory calculations} 
 The calculations for the electronic and magnetic properties of all breathing pyrochlores have been carried out using DFT with the Projector Augmented Wave (PAW) method \cite{ProjectorAugmentedWaveMethod}, as implemented in the Vienna ab initio Simulation Package ~\cite{Kresse1993, Kresse1994, Kresse1996Efficiency, Kresse1996Efficient} (VASP) version 5.4.4. The VASP-supplied PBE~\cite{PBE1996} PAW potentials version 5.4 were used, with the value of the valency of each atomic sphere being 12 for chromium, 14 for iron, 6 for oxygen, 3 for lithium, and 13 for both indium and gallium.
 The PBEsol+U~\cite{PBEsol2008} correction to the exchange and correlation functional is employed to account for the strong electronic correlations on the Cr\textsuperscript{3+} 3$d$ orbitals. By adopting the simplified rotationally invariant formulation of DFT+U \cite{Dudarev1998}, we apply the Hubbard correction \( U_{eff}=U-J \) to address both the on-site interaction strength and the intra-atomic Hund’s rule coupling in the \(\ch{Cr^{3+}}\) 3$d$ orbitals. In this study, the effective Hubbard \( U_{\text{eff}} \) value for Cr is set to 1.0 eV, as it yields the experimentally derived breathing ratio \( J/J' \), as derived from energy mapping analysis on both \(\ch{LiGaCr_4O_8}\) and \(\ch{LiInCr_4O_8}\), as detailed in Fig.~\ref{fig:Ueff_magnetic_exchange_breathing_factor} in the Appendix.
 For the formation energy calculations involving other transition metals, we use Hubbard \(U\) values of 4.0, 3.8, and 3.3 eV for the $d$-electrons on Fe, Mn, and Co, respectively, while the Hund coupling was not considered. These Hubbard values were taken from Ref.~\cite{Jain2011} for Fe, and from Ref.~\cite{Wang2006} for Mn and Co. 
 A plane wave kinetic energy cutoff of 700 eV was used, together with an electronic convergence threshold of 10$^{-6}$ eV and a threshold of maximum force of any one ion of 10$^{-2}$ eV {\AA}$^{-1}$. During relaxation, the unit cell shape, volume, and atomic positions were allowed to fully relax. 
 The Brillouin zone was sampled using a Monkhorst-Pack mesh with a spacing of 0.02 {\AA}$^{-1}$, and the width of the Gaussian smearing set to 0.01 eV.
 In all calculations, we employ spin-polarized DFT with a ferromagnetic arrangement of the spins on the $B$-site for any magnetic cation. This approach is chosen because the antiferromagnetic ordering, which is experimentally identified as the correct state, disrupts the symmetry of the original non-magnetic configuration, transitioning it from cubic to tetragonal symmetry, as experimentally reported in \cite{Lee2016}. 

The formation energy for each substitution attempt of the form LiGa$B_4$O$_8$ and LiIn$B_4$O$_8$ (where \(B=\ch{Cr}, \ch{Fe}, \ch{Mn}\)) was calculated using the following formula, where \(A=\ch{Li}\) and \( A'=\ch{Ga} \) or In:
\begin{multline}
    E_{\text{formation}} = E_{\text{DFT}}[AA'B_4\text{O}_8] -  0.5E_{\text{DFT}} [A_2\text{O}]- \\ 0.5E_{\text{DFT}} [A'_2\text{O}_3]- 2E_{\text{DFT}} [B_2\text{O}_3],
    \label{equation}
\end{multline}
The lattice parameters for each compound were taken from the Materials Project database~\cite{jain2013commentary} to maintain consistency across all phases. For initialization, $A_2$O and $A'_2$O$_3$ were set as non-spin-polarized due to their lack of intrinsic magnetic moments. In contrast, $B_2$O$_3$ was initialized with a spin arrangement that corresponds to the magnetic ground state as specified in the Materials Project. We adopted this setup to ensure that each term in the energy of formation calculation accurately reflects the system's most stable ground-state configuration. 

 For the cation swapping analysis involving \(B\)-site and \(A/A'\)-site ordering, each calculation is performed on the pristine structure, defined as the relaxed reference structure obtained from a spin-polarized DFT calculation with ferromagnetic spin alignment. The process includes swapping the coordinates of the \(B\)-site Cr ion with those of the \(A\)- or \(A'\)-site cation for \(B\)-site ordering analysis, or interchanging coordinates between \(A\)- and \(A'\)-site cations for the \(A/A'\)-site ordering analysis. Each coordinate swap is followed by a full relaxation of the structure, with all spins initially aligned in the up orientation on the magnetic cation, reflecting a ferromagnetic spin arrangement. The ground-state energy is then obtained from this fully relaxed configuration.
The atomic simulation environment was used to manage the calculations \cite{Larsen2017}.

\subsection*{Calculating \(\Delta_{\text{o}}\) using Maximally Localized Wannier Functions}  
The crystal field splitting \(\Delta_{\text{o}}\) for \(\ch{Cr^{3+}}\) in the octahedral \(\ch{CrO_6}\) environment of \(\ch{LiGaCr_4O_8}\) was calculated by evaluating the on-site energies of the \(d\)-orbitals from a tight-binding (TB) model constructed with Maximally Localized Wannier Functions (MLWFs) \cite{Marzari1997}. These Wannier functions were designed to resemble the atomic \(d\)-orbitals of \(\ch{Cr^{3+}}\), allowing the crystal field splitting \(\Delta_{\text{o}}\) to be obtained from the difference in on-site energies of the \(d\)-orbitals of the Cr ions \cite{Scaramucci2014}.
To facilitate the analysis, we performed non-spin-polarized calculation on the primitive cell of \ch{LiGaCr_4O_8}, using the lattice parameters reported in Ref. \cite{Okamoto2013}. A subsequent non-self-consistent field calculation along a high-symmetry path in the Brillouin zone was used to examine the band structure. Projected band analysis showed that Cr \(e_g\) states lie entirely above the Fermi energy, while \(t_{2g}\) states are found both below and above it. To obtain the Wannier functions, calculations were performed with the VASP2WANNIER90 interface \cite{Franchini2012}, focusing on energy bands dominated by \(\ch{Cr^{3+}} \, d\)-states,  spanning from \(-2.84 \, \text{eV}\)  below to \(3.36 \, \text{eV}\) above the Fermi energy. Given the presence of four Cr atoms in the primitive cell, this setup required defining 20 Wannier functions (five \(d\)-orbitals per Cr ion), centered at the \(\ch{Cr}^{3+}\) sites. 

The Wannier90 code \cite{Mostofi2014} was then used to iteratively minimize the spread functional, thereby achieving maximally localized Wannier functions, which are centered on the \(d\)-bands of \(\ch{Cr}^{3+}\). \(\Delta_{\text{o}}\) was determined from the on-site energy difference between \(e_g\) (comprising \(d_{x^2-y^2}\) and \(d_{z^2}\)) and \(t_{2g}\) (comprising \(d_{xy}\), \(d_{yz}\), and \(d_{zx}\)) orbitals, represented as:
\begin{equation}
   \Delta_{\text{o}} = \frac{\left( \epsilon_{d_{x^2-y^2}} + \epsilon_{d_{z^2}} \right)}{2} - \frac{\left( \epsilon_{d_{xy}} + \epsilon_{d_{yz}} + \epsilon_{d_{zx}} \right)}{3}
\end{equation}
where each \(\epsilon\) is the on-site energy obtained from the Wannier-based TB Hamiltonian. Details of these energies, accounting for the lifted orbital degeneracies due to \(\ch{CrO_6}\) octahedral distortion, are provided in Table \ref{table:onsite-energies} in the Appendix.

\subsection*{Energy-mapping method}

The energy-mapping analysis based on four ordered spin states \cite{Xiang2013}, also referred to as the four-state method, is a broken symmetry approach, where broken-symmetry states (instead of eigenstates of exact Hamiltonians) are adopted for energy mapping between the models and results of first-principles calculations \cite{Li2021}.

We use $E_{ij,\alpha \beta}$ ($\alpha, \beta = \uparrow, \downarrow$) to denote the energy of the configuration where spin $i$ is parallel or antiparallel to the $z$ direction (if $\alpha = \uparrow$ or $\downarrow$, respectively), spin $j$ is parallel or antiparallel to the $z$ direction (if $\beta = \uparrow$ or $\downarrow$, respectively), and all the spins except $i$ and $j$ are kept unchanged in the four states.  
Then $J_{ij}$ can be expressed as \cite{Sabani2020}
\begin{equation}
J_{ij} = \frac{E_{ij,\uparrow \uparrow} + E_{ij,\downarrow \downarrow} - E_{ij,\uparrow \downarrow} - E_{ij,\downarrow \uparrow}}{4S^2}
\label{eq:calculation2-spin-hamiltonian}
\end{equation} 
With \(S=\frac{3}{2}\) for \ch{Cr^{3+}}. 

Breathing pyrochlores display a modified spin Hamiltonian, as shown in Equation \ref{eq:BreathingPyroSpinHamiltonian}. This Hamiltonian distinguishes between intra-tetrahedral (\( J \)) and inter-tetrahedral (\( J' \)) magnetic exchange interactions, quantifying the magnetic exchange strength between two Cr atoms within the same smaller and larger tetrahedral units, respectively.

To extract the intra-tetrahedral magnetic exchange interaction, we selected two Cr atoms located at the corners of the smaller tetrahedral unit of the breathing pyrochlore structure as reference atoms. The remaining spins in the conventional unit cell of \(\ch{LiGaCr_4O_8}\) and \(\ch{LiInCr_4O_8}\) were kept constant to isolate the effect of the reference spins. We determined the intra-tetrahedral exchange parameter \( J\) using the ground-state energies \( E_{ij,\uparrow \uparrow} \), \( E_{ij,\uparrow \downarrow} \), \( E_{ij,\downarrow \uparrow} \), and \( E_{ij,\downarrow \downarrow} \), which correspond to the four possible spin arrangements for the two reference Cr atoms. Similarly, to determine the inter-tetrahedral exchange parameter \( J' \), we selected two Cr atoms within the larger tetrahedral unit as reference atoms and applied the same approach. This two-step process allowed us to derive the isotropic magnetic exchange parameters, with the values provided in Table \ref{table:exchange-constants}. For a comprehensive overview, Table \ref{tab:supplement-material-energies-energymapping} lists the ground-state energies associated with each specific spin configuration used in the energy mapping method to calculate these exchange interactions, to which the interested reader is referred for further details.

\section*{Data availability}
The data that support the findings of this study are available from the corresponding author upon reasonable request.

\begin{acknowledgments}
This work was supported by the Natural Sciences and Engineering Research Council of Canada (NSERC), the Canadian Institute for Advanced Research (CIFAR), and the Sloan Research Fellowships program. VM was supported by Quantum Pathways and UBC's Work Learn International Undergraduate Research Award. This research was undertaken thanks in part to funding from the Canada First Research Excellence Fund, Quantum Materials and Future Technologies Program.
\end{acknowledgments}

\section*{Author contributions}

This study was initiated and supervised by K.M.K. and A.M.H. Solid state synthesis and x-ray diffraction was performed by J.M., A.A.M.S., and M.R.R. DFT and OSPE calculations were performed by V.M. and S.S.A. with guidance from J.R. The manuscript was written by V.M., S.S.A., A.A.M.S., and A.M.H. with input from all authors.

\section*{Appendix}
\label{sec:Supplementary}

{
\renewcommand{\belowrulesep}{0pt}
\renewcommand{\aboverulesep}{0pt}
{
\setlength{\tabcolsep}{0pt}
\begin{table}[H]

\resizebox{\columnwidth}{!}{
\begin{tabular}{lcccccc}
\toprule
\rowcolor[HTML]{DADADA} 
\multicolumn{1}{c}{\cellcolor[HTML]{DADADA}} &
  \cellcolor[HTML]{DADADA} &
  \multicolumn{3}{c}{\cellcolor[HTML]{DADADA}\textbf{Coordinates}} &
  \cellcolor[HTML]{DADADA} &
  \cellcolor[HTML]{DADADA} \\
\rowcolor[HTML]{DADADA} 
\multicolumn{1}{c}{\multirow{-2}{*}{\cellcolor[HTML]{DADADA}\textbf{ Atom }}} &
  \multirow{-2}{*}{\cellcolor[HTML]{DADADA}\textbf{ Wyck. }} &
  \textbf{x} &
  \textbf{y} &
  \textbf{z} &
  \multirow{-2}{*}{\cellcolor[HTML]{DADADA}\textbf{Occ.  }} &
  \multirow{-2}{*}{\cellcolor[HTML]{DADADA}\textbf{\( \mathbf{B_{eq}}  \)(\AA \(\mathbf{^2}\)) \hspace{1pt} }} \\ \hline
Li  & 4a  & 0         & 0          & 0          & 1 & 3(1)    \\
\rowcolor[HTML]{D5D5D5} 
Ga  & 4d  & 0.75      & 0.75       & 0.75       & 1 & 2.2(1)  \\
Cr  & 16e \hspace{2pt}&  0.3715(4) \hspace{1pt} & 0.37158(4) \hspace{1pt} &  0.37158(4) & 1 & 1.57(7) \\
\rowcolor[HTML]{DADADA} 
O1 & 16e \hspace{2pt}&  0.1346(7) \hspace{1pt} &  0.1346(7) \hspace{1pt}  &  0.1346(7)  & 1 & 1.3(3)  \\
O2 & 16e \hspace{2pt}&  0.6177(9) \hspace{1pt} &  0.6177(9) \hspace{1pt}  &  0.6177(9)  & 1 & 1.6(3)  \\ \bottomrule
\end{tabular}

}

\caption{Refined crystal structure for \ch{LiGaCr_4O_8} in the $F\overline{4}3m$ space group with cubic lattice parameter $a = $~8.2443(1)~\AA\ including Wyckoff site (Wyck.), atomic coordinates, occupancies (Occ.) and thermal parameters ($B_\text{eq}$). Standard errors are represented by the number in parenthesis on the final digit. Values without errors were not refined.}
\label{tab:LiGaCr4O8refinement}
\end{table}

}

{
\setlength{\tabcolsep}{0pt}
\begin{table}[H]

\centering
\resizebox{\columnwidth}{!}{
\begin{tabular}{lcccccc}
\toprule
\rowcolor[HTML]{DADADA} 
\multicolumn{1}{c}{\cellcolor[HTML]{DADADA}} &
  \cellcolor[HTML]{DADADA} &
  \multicolumn{3}{c}{\cellcolor[HTML]{DADADA}\textbf{Coordinates}} &
  \cellcolor[HTML]{DADADA} &
  \cellcolor[HTML]{DADADA} \\
\rowcolor[HTML]{DADADA} 
\multicolumn{1}{c}{\multirow{-2}{*}{\cellcolor[HTML]{DADADA}\textbf{ Atom }}} &
  \multirow{-2}{*}{\cellcolor[HTML]{DADADA}\textbf{ Wyck. }} &
  \textbf{x} &
  \textbf{y} &
  \textbf{z} &
  \multirow{-2}{*}{\cellcolor[HTML]{DADADA}\textbf{Occ. }} &
  \multirow{-2}{*}{\cellcolor[HTML]{DADADA}\textbf{\( \mathbf{B_{eq}}  \)(\AA \(\mathbf{^2}\)) \hspace{1pt} } } \\ \hline
Li1 &                      &                             &                             &                             & 0.90(1)  & 3(1)    \\
Fe1 & \multirow{-2}{*}{4b} & \multirow{-2}{*}{0.625}     & \multirow{-2}{*}{0.625}     & \multirow{-2}{*}{0.625}     & 0.09(1)  & 3(1)    \\
\rowcolor[HTML]{DADADA} 
Fe2 &                      &                             &                             &                             & 0.768(4) & 0.23(9) \\
\rowcolor[HTML]{DADADA} 
Ga1 & 12d                  & 0.875                       & 0.8665(2)                   & 0.6165(2)                   & 0.2      & 0.23(9) \\
\rowcolor[HTML]{DADADA} 
Li2  &                      &                             &                             &                             & 0.031(4) & 0.23(9) \\
Fe3  &                      &                             &                             &                             & 0.8      & 0.27(9) \\
Ga2  & \multirow{-2}{*}{8c} & \multirow{-2}{*}{0.2528(3)} & \multirow{-2}{*}{ 0.2528(3)} & \multirow{-2}{*}{ 0.2528(3)} & 0.2      & 0.27(9) \\
\rowcolor[HTML]{DADADA} 
O1 & 8c                   & 0.382(1)                    & 0.382(1)                    & 0.382(1)                    & 1        & 1.1(3)  \\
O2 & 24e                  & 0.122(1)                    & 0.128(1)                    & 0.379(1)                    & 1        & 0.5(2) \\ \bottomrule
\end{tabular}
}

\caption{Refined crystal structure for \(\ch{Li(Ga_{0.2}Fe_{0.8})_5O_8}\) in the $P4_332$ space group with cubic lattice parameter $a = $~8.3107(1)~\AA\ including Wyckoff site (Wyck.), atomic coordinates, occupancies (Occ.) and thermal parameters ($B_\text{eq}$). Standard errors are represented by the number in parenthesis on the final digit. Values without errors were not refined.}
\label{tab:LiGaFe4O8refinement}
\end{table}
}
}

\begin{table}[H]
\centering
\begin{tabular}{lcccc}
\toprule
\multicolumn{5}{c}{\textbf{\ch{LiGaCr_4O_8}}} \\
\midrule
 & \makecell{\bm{$d$} \\ \textbf{(\AA)}}  & \makecell{\bm{$d'$} \\ \textbf{(\AA)}} & \makecell{{$\sfrac{d}{d'}$} \\ \textbf{ }}  & \makecell{\textbf{Cr-O-Cr} \\ \textbf{(°)}} \\
\midrule
Experimental \cite{Okamoto2013} & 2.86204 & 2.96459 & 0.965 & 94.1715 \\
Spin-polarized & 2.85682 &	2.99089 &	0.955 &	94.4432 \\
Non-spin-polarized & 2.44253 & 3.20967 & 0.761	& 81.1925 \\
\midrule
 \multicolumn{5}{c}{\textbf{\ch{LiInCr_4O_8}}} \\
\midrule
 & \makecell{\bm{$d$} \\ \textbf{(\AA)}}  & \makecell{\bm{$d'$} \\ \textbf{(\AA)}} & \makecell{{$\sfrac{d}{d'}$} \\ \textbf{ }} & \makecell{\textbf{Cr-O-Cr} \\ \textbf{(°)}} \\
\midrule
Experimental \cite{Okamoto2013} & 2.903 & 3.051 & 0.951 & 94.5904 \\
Spin-polarized & 2.90941 & 3.058 &	0.951 &	95.0512\\
Non-spin-polarized & 2.48903 & 3.267 & 0.7612 & 81.8504 \\
\bottomrule
\end{tabular}
\caption{Comparison of small tetrahedral distance ($d$), large tetrahedral distance ($d'$), their ratio ($\sfrac{d}{d'}$), and the Cr-O-Cr angle for \ch{LiGaCr_4O_8} and \ch{LiInCr_4O_8} calculated with spin-polarized, and non-spin-polarized DFT and compared to experimental values.}
\label{table:supplementaryInformation-breathingpyrochlore-distortion}
\end{table}

\begin{table}[H]
\centering
\begin{tabular*}{\columnwidth}{@{\extracolsep{\stretch{1}}}*{2}{lccc}@{}}
\toprule
\textbf{Orbital} & \textbf{Energy (eV)} \\
\midrule
$\epsilon_{d_{xy}}$ & 6.08183 \\
$\epsilon_{d_{yz}}$ & 6.08183 \\
$\epsilon_{d_{zx}}$ & 6.08153 \\
\midrule  
\textbf{Mean ($E_{t_{2g}}$)} & \(6.08173 \pm 0.00008 \)\\
\midrule

$\epsilon_{d_{x^2-y^2}}$ & 8.32930 \\
$\epsilon_{d_{z^2}}$ & 8.32935 \\
\midrule
\textbf{Mean ($E_{e_{g}}$)} & $8.32932  \pm 0.00002$ \\
\midrule
\textbf{$\Delta_{\text{o}}$} & $2.2476 \pm 0.0001 $\\ 
\midrule
\textbf{OSPE (-$\frac{38}{45}\Delta_{\text{o}}$)} & -$1.8980 \pm 0.0001$\\ 
\bottomrule
\end{tabular*}
\caption{On-site energies of \ch{Cr^{3+}}-centered \(d\)-like Wannier functions, along with their averages and associated uncertainties for the \(t_{2g}\) (3-fold degeneracy) and \(e_g\) (2-fold degeneracy) orbitals. The uncertainties for the \(t_{2g}\) and \(e_g\) mean values were calculated based on the standard deviation divided by the square root of the sample size. The crystal field splitting \(\Delta_{\text{o}}\) is determined as the difference between the mean on-site energies of \(E_{t_{2g}}\) and \(E_{e_g}\), while the octahedral site preference energy (OSPE) is calculated as \(\frac{38}{45}\Delta_{\text{o}}\). }
\label{table:onsite-energies}
\end{table}

\begin{table}[H]
\centering
\begin{tabular*}{\columnwidth}{@{\extracolsep{\stretch{1}}}cccc}
\toprule
\multicolumn{3}{c}{\textbf{\ch{LiGaCr_4O_8}}} \\
\midrule
 & \textbf{Energy (eV)}& \textbf{Energy (eV)} \\
\midrule
\( E_A =  E_{ij,\downarrow \downarrow} \)  &-433.00807  & -432.91058  \\
\( E_B = E_{ij,\downarrow \uparrow}  \) & -433.02794 & -433.00807 \\
\( E_C = E_{ij,\uparrow \downarrow}  \) & -433.03587& -433.00807 \\
\( E_D = E_{ij,\uparrow \uparrow}\)  & -433.00808 & -433.03587 \\
\midrule
 & \textbf{Inter-tet. (\bm{$J'$})} & \textbf{Intra-tet. (\bm{$J$})} \\
\midrule
\( J_{ij} \) (eV) & 0.0053  & 0.00774 \\
 \midrule
$B_f=\sfrac{J'}{J}$ & \multicolumn{2}{c}{0.68475}\\
\toprule
\multicolumn{3}{c}{\textbf{\ch{LiInCr_4O_8}}} \\
\midrule
  & \textbf{Energy (eV)}& \textbf{Energy (eV)}\\
\midrule
\( E_A =  E_{ij,\uparrow \downarrow} \)  &  -423.8030221 & -423.8140159 \\
\( E_B = E_{ij,\downarrow \uparrow}  \) & -423.8029053 & -423.8171094 \\
\( E_C = E_{ij,\downarrow \downarrow}  \) & -423.7941603 & -423.8030679 \\
\( E_D = E_{ij,\uparrow \uparrow}\)  & -423.8140205 & -423.8030641 \\
\midrule
 & \textbf{Inter-tet. (\bm{$J'$})} & \textbf{Intra-tet. (\bm{$J$})} \\
\midrule
\( J_{ij} \) (eV) & 0.000250382 & 0.002777036 \\
\midrule
\(B_f=\sfrac{J'}{J}\) & \multicolumn{2}{c}{0.09016}\\
\bottomrule
\end{tabular*}
\caption{Ground-state energies of four distinct spin arrangements on the Cr atoms in the conventional unit cell of \ch{LiGaCr_4O_8} and \ch{LiInCr_4O_8}, along with the nearest-neighbor magnetic exchange interactions calculated using Equation \ref{eq:calculation2-spin-hamiltonian} and the breathing ratio \(B_f=\sfrac{J'}{J}\). }
\label{tab:supplement-material-energies-energymapping}
\end{table}

\begin{figure}[H]
\includegraphics[width=\columnwidth]{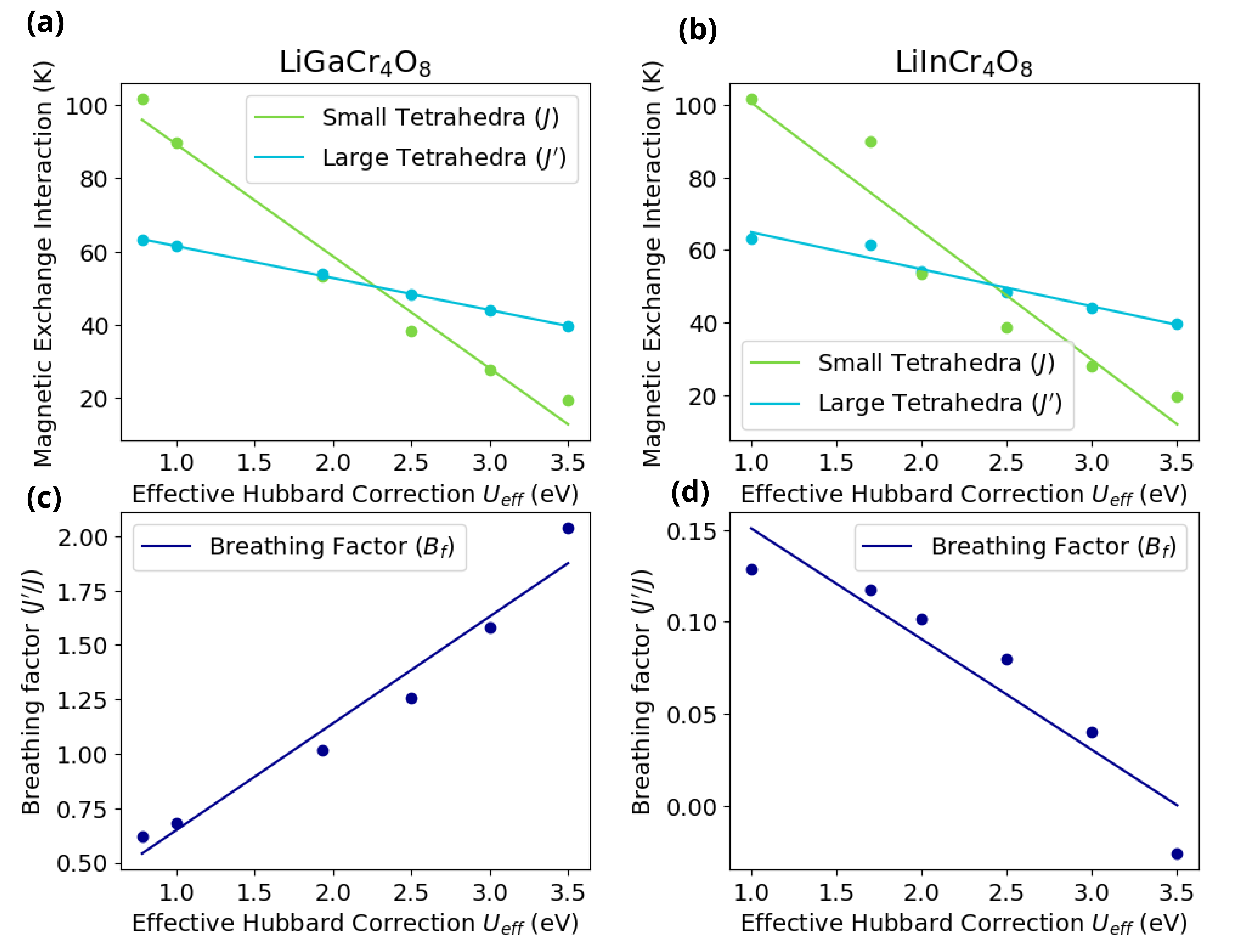}
\caption{\textbf{Influence of the effective Hubbard correction (\( U_{\text{eff}} \)) on the magnetic exchange interactions and the breathing factor for \(\ch{LiGaCr_4O_8}\) and \(\ch{LiInCr_4O_8}\).} The variation in magnetic exchange interactions \(J\) and \(J'\) (in K) within the small and large tetrahedra, respectively, as a function of \( U_{\text{eff}} \) for \textbf{(a)} \(\ch{LiGaCr_4O_8}\) and \textbf{(c)} \(\ch{LiInCr_4O_8}\). The exchange interactions decrease with increasing \( U_{\text{eff}} \) for both materials, although \(\ch{LiInCr_4O_8}\) shows a non-monotonic trend.}
\label{fig:Ueff_magnetic_exchange_breathing_factor}
\end{figure}

\newpage
\bibliography{bibliography}

\end{document}